\documentclass[%
 aip,
 jcp,
 amsmath,amssymb,
preprint,%
]{revtex4-2}

\usepackage{graphicx}
\usepackage{mhchem}
\usepackage{dcolumn}
\usepackage{bm}
\usepackage[utf8]{inputenc}
\usepackage[T1]{fontenc}
\usepackage{etoolbox}
\usepackage{booktabs}
\usepackage{hyperref}
\usepackage{xcolor}
\usepackage{physics}

\usepackage[normalem]{ulem}

\newcommand{\cre}[1]{a_{#1}^{\dag}}
\newcommand{\ani}[1]{a_{#1}^{}}

\newcommand{\PsiCC}{\mathrm{CC}}
\newcommand{\bx}{\boldsymbol{x}}

\newcommand{\adafqmc}{\textsc{ad-afqmc}}

\makeatletter
\def\@email#1#2{%
 \endgroup
 \patchcmd{\titleblock@produce}
  {\frontmatter@RRAPformat}
  {\frontmatter@RRAPformat{\produce@RRAP{*#1\href{mailto:#2}{#2}}}\frontmatter@RRAPformat}
  {}{}
}%
\makeatother
\begin{document}

\title{Systematic improvement of trial states in phaseless auxiliary-field quantum Monte Carlo}
\author{Eirik F.~Kjønstad}
\affiliation{Division of Chemistry and Chemical Engineering, California Institute of Technology, Pasadena, California 91125, USA}
\author{Yann Damour}
\affiliation{Division of Chemistry and Chemical Engineering, California Institute of Technology, Pasadena, California 91125, USA}
\author{Sandeep Sharma}
\affiliation{Division of Chemistry and Chemical Engineering, California Institute of Technology, Pasadena, California 91125, USA}
\author{Garnet Kin-Lic Chan$^\ast$}
\email{gkc1000@gmail.com}
\affiliation{Division of Chemistry and Chemical Engineering, California Institute of Technology, Pasadena, California 91125, USA}

\begin{abstract}
We extend the use of  coupled cluster (CC) trial states in the phaseless auxiliary-field quantum Monte Carlo (AFQMC) method beyond single and double excitations to include both triple and quadruple excitations. With this AFQMC/CC hierarchy, we are able to systematically benchmark the method's performance on molecular systems as the quality of the trial is improved. Our results show that 
the phaseless AFQMC energy 
improves systematically and is typically significantly more accurate than the energy of the underlying trial state. However, the relative improvement compared to the trial CC energy decreases as we ascend the CC hierarchy. As the CC wavefunction is usually further approximated when used as an AFQMC trial, we also explore the relationship between the components of the CC wavefunction and the resulting AFQMC/CC error. Our results suggest that improving the representation of the CC wave function in the AFQMC trial does not always lower the bias even when it increases the fidelity of the trial with the exact ground state. 
\end{abstract}

\maketitle

\section{Introduction}
The phaseless auxiliary-field quantum Monte Carlo (AFQMC) method\citep{zhang2003quantum} provides a path to determine the ground state of the electronic Schrödinger equation via stochastic imaginary time propagation of an ensemble of non-orthogonal Slater determinants.
To avoid the fermion sign problem and its associated exponential cost, the method imposes a phaseless constraint on the propagation,
introducing, in the process, a bias whose magnitude depends on an approximation of the ground state referred to as a trial. 
In a variety of chemical benchmarks, it has been found that
for the simple case of a mean-field trial state, the bias generated in the phaseless AFQMC method yields an accuracy\citep{motta2018ab,lee2022twenty} slightly better than the coupled cluster (CC)\citep{bartlett2007coupled} method at the level of single and double (CCSD)\citep{purvis1982full} excitations. 

As the trial improves, the error in the energy typically shrinks, and an exact trial trivially recovers the exact energy due to the form of the energy estimator.
However, the relationship between the trial and the resulting phaseless constraint bias remains poorly understood, although some recent works have attempted to benchmark both the standard and alternative constraints.\citep{sukurma2023benchmark,weber2023expanding} Despite this fact, a variety of approaches to obtain more accurate trial states have been pursued over the past two decades, including mean-field trials,\citep{al2006auxiliary,purwanto2008eliminating} matrix product states,\citep{wouters2014projector,jiang2025unbiasing} and a number of strategies to determine compact multi-Slater determinant (MSD) expansions to serve as the trial state.\citep{purwanto2009excited,landinez2019non,mahajan2022selected,sukurma2025self} 
In a distinct but related line of research, preparing the trial state on a quantum computer has also been pursued in recent work.\citep{huggins2022unbiasing,amsler2023classical,kiser2024classical,huang2024evaluating,kiser2025contextual,jiang2025walking}
Overall, these efforts have shown that it is possible in practice to improve upon the bias of single-determinant trial states, but it remains challenging to reach chemical accuracy at moderate computational cost for a broad class of chemical problems.

One motivating assumption behind this work is that there
exist more accurate trial states for which the phaseless AFQMC method gives 
improved
estimates at 
a lower effort than that needed to refine the estimates by other means. This is indeed plausible, given the high accuracy and favorable $N^5$
scaling of the method for mean-field trials (for fixed stochastic error, where $N$ is system size). However, 
the question remains of how much one should expect a given controlled improvement of the trial to translate into a corresponding improvement in the AFQMC estimates.

In the present work, we address this question by benchmarking phaseless AFQMC using a hierarchy of CC trial states, obtained by extending the recent CCSD implementation by Mahajan \emph{et al.}\citep{mahajan2025beyond} to support trials with triple (CCSDT)\citep{noga1987full} and quadruple (CCSDTQ)\citep{kucharski1992coupled} excitations. In Ref.~\citenum{mahajan2025beyond}, using a CCSD trial state was found to outperform the approximate triples method CCSD(T),\citep{raghavachari1989fifth} 
suggesting that an AFQMC/CC hierarchy might be a particularly efficient route to the exact ground-state energy. 

The paper is organized as follows.
In Section \ref{sec:theory} we describe the phaseless AFQMC method and our implementation approach for trial states with triple and quadruple excitations. This is followed in Section \ref{sec:results-discussion} by a presentation and discussion of our benchmark results and in Section \ref{sec:conclusions} by concluding remarks. 

\section{Theory} \label{sec:theory}

\subsection{Phaseless auxiliary field quantum Monte Carlo}
The ground state $\ket{\psi_0}$ of a Hamiltonian $H$ can be determined by propagating in imaginary time $\tau$ any guess $\ket{\phi}$ that has a non-zero overlap with $\ket{\psi_0}$, i.e.,
\begin{align}
    \ket{\psi_0} \propto \lim_{\tau \rightarrow \infty} U(\tau) \ket{\phi} = \lim_{\tau \rightarrow \infty} e^{-\tau H} \ket{\phi}.
\end{align}
In the AFQMC method, the propagator $U(\Delta \tau)$ is decomposed into an integral over a set of auxiliary fields $\bx = \qty{x_\gamma}$ associated with one-body interactions $B(\bx)$,
\begin{align}
    U(\Delta \tau) = e^{-\Delta \tau H} = \int\dd\bx \, P(\bx) B(\bx) + \mathcal{O}(\Delta \tau^2). \label{eq:short-time-no-details}
\end{align}
To obtain a representation of $\ket{\psi_0}$, one then propagates an ensemble of non-orthogonal Slater determinants (walkers) by sampling field configurations $\bx$ from $P(\bx)$ and applying $B(\bx)$ to the walkers. 

These auxiliary fields are introduced via a Hubbard-Stratonovich (HS) transformation of the two-body terms in the Hamiltonian.\citep{hubbard1959,stratonovich1957} 
Let
\begin{align}
    H = \sum_{pq\sigma} h_{pq} \cre{p\sigma} \ani{q\sigma} + \frac{1}{2} \sum_{pqrs\sigma\tau} g_{prqs} \cre{p\sigma} \cre{q\tau} \ani{s\tau} \ani{r\sigma},
\end{align}
where $p,q,r,s$ denote molecular orbitals and $\sigma, \tau$ denote spin coordinates. By applying a Cholesky decomposition of
the two-body integrals, i.e., 
    $g_{prqs} = \sum_{\gamma} L_{pr}^\gamma L_{qs}^\gamma$,
then $H$ can be recast as
\begin{align}
    H = \bar{h} - \frac{1}{2} \sum_\gamma v_\gamma^2,
\end{align}
with 
\begin{align}
    \bar{h} = \sum_{pq\sigma} \bar{h}_{pq} \cre{p\sigma} \ani{q\sigma}, \quad v_\gamma = i \sum_{pq\sigma} L^\gamma_{pq} \cre{p\sigma} \ani{q\sigma},
\end{align}
and $\bar{h}_{pq} = h_{pq} - \frac{1}{2} \sum_r g_{prrq}$. 
The two-body propagator is then decomposed via the HS transformation, giving 
\begin{align}
\begin{split}
    U(\Delta \tau) 
    &\approx e^{-\frac{\Delta \tau}{2} \bar{h}} \prod_\gamma \int \frac{\dd x_\gamma}{\sqrt{2 \pi}}e^{-\frac{x_\gamma^2}{2}} e^{\sqrt{\Delta \tau} x_\gamma v_\gamma} e^{-\frac{\Delta \tau}{2} \bar{h}},
\end{split}
\end{align}
which is of the form given in Eq.~\eqref{eq:short-time-no-details}. By sampling a field configuration $\bx$, we obtain a short-time propagator $B(\bx)$ that transforms a walker into 
another walker. In the importance sampling representation, 
the ensemble of walkers is 
\begin{align}
    \ket{\psi} = \sum_k w_k \frac{\ket{\phi_k}}{\braket{\psi_T}{\phi_k}},
\end{align}
where 
$w_k$ are weights that are updated in each time step based on an importance sampling obtained from a 
trial state $\ket{\psi_T}$. In particular, in the ``hybrid'' formalism, the weight of the walker $\vert \phi_k \rangle$ is updated by multiplication with the importance function\citep{motta2018ab}
\begin{align}
    I_k = \frac{\langle \psi_T \vert B(\boldsymbol{x}-\bar{\boldsymbol{x}}) \vert \phi_k \rangle}{\langle \psi_T \vert \phi_k \rangle} e^{\boldsymbol{x} \cdot \bar{\boldsymbol{x}} - \frac{1}{2} \bar{\boldsymbol{x}}\cdot\bar{\boldsymbol{x}}},
\end{align}
where $\bar{\boldsymbol{x}}$ is a shift in the auxiliary fields that stabilizes the propagation. This shift is given as
\begin{align}
    \bar{x}_\gamma = - \sqrt{\Delta \tau} \; v^L_\gamma(\phi_k) =  - \sqrt{\Delta \tau} \frac{\langle \psi_T \vert v_\gamma \vert \phi_k \rangle}{\langle \psi_T \vert \phi_k \rangle}, \label{eq:force-bias}
\end{align}
where we have introduced the local mixed estimator 
\begin{align}
    O_L(\phi_k) = \frac{\langle \psi_T \vert O \vert \phi_k \rangle}{\langle \psi_T \vert \phi_k \rangle}.
\end{align}
The ground state energy is evaluated from the mixed estimator
\begin{align}
    E = \langle H \rangle_m = \frac{\langle \psi_T \vert H \vert \psi \rangle}{\langle \psi_T \vert \psi \rangle} = \frac{\sum_k w_k H_L(\phi_k)}{\sum_k w_k} \equiv \frac{\sum_k w_k E_L(\phi_k)}{\sum_k w_k}. \label{eq:energy-mixed-estimator} 
\end{align}

For an arbitrary Hamiltonian, the propagator $B(\mathbf{x})$ gives rise to complex weights. Without any constraints, the AFQMC method then suffers from the sign problem because the phases of the walkers become uniformly distributed in the complex plane, leading to a formal exponential computational scaling for a fixed stochastic error.
This problem can be avoided at the cost of introducing a bias that depends on the trial. In the \emph{phaseless} approximation, $\ket{\psi_T}$ is used to redefine the 
weights by projecting the importance function onto the positive real axis, i.e.,\citep{zhang2003quantum,motta2018ab}
\begin{align}
    w_k \mapsto w_k \cdot \abs{I_k} \cdot \max(0, \cos \theta_k),
\end{align}
where $\theta_k = \arg I_k$. 

\subsection{Trial-specific implementation aspects}
The implementation for a specific type of trial state involves evaluation of the overlap with an arbitrary Slater determinant $\vert \phi \rangle$,
\begin{align}
    O(\phi) = \braket{\psi_T}{\phi},
\end{align}
as well as the local energy associated with $\ket{\phi}$,
\begin{align}
    E_L(\phi) = \frac{\mel*{\psi_T}{H}{\phi}}{\braket{\psi_T}{\phi}} 
    ,
\end{align}
which enters in the mixed energy estimator in Eq.~\eqref{eq:energy-mixed-estimator}.
The one-body contribution to the local energy is also used to evaluate the force bias $\bar{x}_\gamma $, see Eq.~\eqref{eq:force-bias}.

Assuming $\vert \psi_T \rangle$ can be written as ${C}|\phi_0\rangle$, where ${C}$ is an operator with a compact expansion in second quantized operators and $|\phi_0\rangle$ is a reference determinant, both  $O(\phi)$ and $E_L(\phi)$ can be evaluated efficiently by 
applying the generalized Wick's theorem.\citep{balian1969nonunitary} 
The resulting expressions consist of contractions between coefficients of ${C}$, the integrals in $H$, and products of Green's functions, defined as
\begin{align}
    G_{p\sigma,q\tau} = \frac{\mel*{\phi_0}{\cre{p\sigma} \ani{q\tau}}{\phi}}{\braket{\phi_0}{\phi}} = \delta_{\sigma\tau} (U_\sigma(V_\sigma^\dagger U_\sigma)^{-1}V_\sigma^\dagger)_{qp},
\end{align}
where $\braket{\phi_0}{\phi} = \prod_\sigma \det (V_\sigma^\dagger U_\sigma)$ and $U_\sigma$ and $V_\sigma$ denote the orbital coefficients of $\ket{\phi}$ and $\ket{\phi_0}$.

The implementation of $E_L(\phi)$, while 
straightforward, can be rather complicated for some trial wave functions. A convenient approach is therefore to instead implement it from derivatives of $O(\phi)$.\citep{filippi2016simple,mahajan2025beyond} In particular, if we denote
\begin{align}
    O(\phi,\lambda; X) = \mel*{\psi_T}{e^{\lambda X}}{\phi},
\end{align}
then 
\begin{align}
    E_L(\phi) = \dv{\lambda} O(\phi,\lambda; \bar{h})\big\vert_0 + \frac{1}{2} \sum_\gamma \dv[2]{\lambda_\gamma} O(\phi,\lambda_\gamma; v_\gamma)\big\vert_0.
    \label{eq:ovlp_derivative}
\end{align}
Note that since $e^{\lambda X}$ is a one-body rotation, $e^{\lambda X}\vert \phi \rangle$ is a Slater determinant, implying that the overlap derivatives in Eq.~\eqref{eq:ovlp_derivative} can be evaluated using an implementation of $O(\phi)$ for an arbitrary Slater determinant $\ket{\phi}$. 

\subsection{Coupled cluster trials} \label{sec:ccsdt-overlap-impl}
Mahajan \emph{et al.}\citep{mahajan2025beyond} applied automatic differentiation to evaluate 
Eq.~\eqref{eq:ovlp_derivative} for CCSD trials projected onto the corresponding CI subspace (CISD). 
In this work, we extend  
the implementation to trials that incorporate triple and quadruple excitations. The CC ground state can be expressed as an exponential ansatz,
\begin{align}
    \ket{\PsiCC} = e^T \ket{\phi_0} = \qty(1 + T + \frac{1}{2}T^2 + \ldots) \ket{\phi_0},
\end{align}
where $T = \sum_\mu t_\mu \tau_\mu = \sum_{i=1}^n T_i$ incorporates excitation operators ($\tau_\mu$) up to some excitation order $n$. The reference $\ket{\phi_0}$ is typically a Hartree-Fock determinant, and the amplitudes ($t_\mu$) are determined by projection of the Schrödinger equation.\citep{bartlett2007coupled}

Due to its exponential form, $\ket{\PsiCC}$ can only be used as a trial in small systems since the evaluation of $O(\phi)$ quickly becomes intractable. This can be overcome by projection onto a CI subspace, i.e.,\citep{mahajan2025beyond}
\begin{align}
    \ket{\psi_T} = \mathcal{P} \ket{\PsiCC}, \label{eq:projected-cc-trial}
\end{align}
with the projection operator
\begin{align}
    \mathcal{P} = \ketbra{\phi_0} + \sum_\mu \ketbra{\mu} = \sum_{i=0}^k \mathcal{P}_i.
\end{align}
The simplest choice is to project onto the same subspace as the $T$ truncation, i.e.~$n = k$ (e.g., this means projecting the CCSD state into a CISD state). However, one can also project onto a larger subspace ($n < k$) to include higher-order disconnected cluster contributions in the trial, although this generally increases the computational costs. 

To implement $O(\phi)$ for the trial in Eq.~\eqref{eq:projected-cc-trial}, we determine its effective CI expansion, i.e.,
\begin{align}
    \ket{\psi_T} = \mathcal{P} \ket{\PsiCC} = (1 + C_1 + C_2 + \ldots + C_k) \ket{\phi_0},
\end{align}
where
\begin{align}
\begin{split}
    C_1 &= T_1 \\
    C_2 &= T_2 + \frac{T_1^2}{2} \\
    C_3 &= T_3 + T_1 T_2 + \frac{T_1^3}{6} \\
    &\vdots
\end{split} \label{eq:C-from-T}
\end{align}

\subsubsection*{Expressions for trials with triple and quadruple excitations}
Below we derive the contributions to $O(\phi)$ arising from triples excitations in the spin-unrestricted case. Contributions arising from quadruple excitations are derived analogously and are provided in Appendix \ref{app:ccsdtq}. For contributions to $O(\phi)$ arising from the reference, single, and double excitations, we refer to Ref.~\citenum{mahajan2025beyond}. 

In the spin-orbital basis,\citep{shavitt2009many}
\begin{align}
\begin{split}
    T_1 &= \sum_{ai} t_i^a \cre{a} \ani{i} \\
    T_2 &= \frac{1}{(2!)^2} \sum_{abij} t_{ij}^{ab} \cre{a} \cre{b} \ani{j} \ani{i} \\
    T_3 &= \frac{1}{(3!)^2} \sum_{abcijk} t_{ijk}^{abc} \cre{a} \cre{b} \cre{c} \ani{k} \ani{j} \ani{i}, 
\end{split} 
\end{align}
which we can express in the spin-unrestricted basis as
\begin{align}
\begin{split}
    T_1 &= T_1^\alpha + T_1^\beta \\
    T_2 &= \frac{1}{4} T_2^{\alpha\alpha} + T_2^{\alpha\beta} + \frac{1}{4} T_2^{\beta\beta} \\
    T_3 &= \frac{1}{36} T_3^{\alpha\alpha\alpha} + \frac{1}{4} T_3^{\alpha\alpha\beta} + \frac{1}{4} T_3^{\alpha\beta\beta} + \frac{1}{36} T_3^{\beta\beta\beta} 
\end{split} \label{eq:T-unrestricted}
\end{align}
where, e.g., $T_2^{\alpha\beta}$ consists of double excitations with mixed $\alpha\beta$ excitations, i.e., $\cre{a\alpha} \cre{b\beta} \ani{j\beta} \ani{i\alpha}$. The pre-factor for each spin block is obtained by collecting redundant excitations and is given as $\binom{n}{n-n_\alpha}^2/(n!)^2$ for $T_n^{s}$, where $n_\alpha$ denotes the number of $\alpha$ excitations in the spin block $s$. 

We substitute Eq.~\eqref{eq:T-unrestricted} into Eq.~\eqref{eq:C-from-T} to obtain the expressions for $C_3$:
\begin{align}
    C_3 = \frac{1}{36} C_3^{\alpha\alpha\alpha} + \frac{1}{4} C_3^{\alpha\alpha\beta} + \frac{1}{4} C_3^{\alpha\beta\beta} + \frac{1}{36} C_3^{\beta\beta\beta},
\end{align}
where
\begin{align}
\begin{split}
    C_3^{\alpha\alpha\alpha} &= T_3^{\alpha\alpha\alpha} \\
    C_3^{\alpha\alpha\beta} &= T_3^{\alpha\alpha\beta} + T_1^\beta T_2^{\alpha\alpha} + 4 \, T_1^\alpha T_2^{\alpha\beta} + 2 \, (T_1^\alpha)^2 T_1^\beta \\
    C_3^{\alpha\beta\beta} &= T_3^{\alpha\beta\beta} + T_1^\alpha T_2^{\beta\beta}  + 4 \, T_1^\beta T_2^{\alpha\beta} + 2 \, (T_1^\beta)^2 T_1^\alpha \\
    C_3^{\beta\beta\beta} &= T_3^{\beta\beta\beta}.
\end{split} \label{eq:C3-spin-blocks}
\end{align}

After forming the blocks in Eq.~\eqref{eq:C3-spin-blocks}, we symmetrize same-spin indices in the $C_3$ tensors with respect to exchange of excitations ($ai \leftrightarrow bj$) and anti-symmetrize them with respect to exchange of occupied ($i \leftrightarrow j$) and virtual ($a \leftrightarrow b$) orbitals. This implies no loss of generality and follows from the anti-commutation relations of the fermionic operators.

Next, we define the scaled overlap
\begin{align}
\begin{split}
    \tilde{O}(\phi) = \frac{O(\phi)}{O_0(\phi)} &= \frac{\mel*{\phi_0}{(1 + C_1 + \ldots + C_k)}{\phi}}{\braket{\phi_0}{\phi}} \\
    &= \sum_{i = 0}^k \tilde{O}_i(\phi) =  \sum_{i = 0}^k \sum_{s_i} \tilde{O}_i^{s_i}(\phi),
\end{split}
\end{align}
where $s_i$ denotes the spin blocks in $C_i$. With this notation, 
\begin{align}
\begin{split}
    \tilde{O}_3(\phi) &= \tilde{O}_3^{\alpha\alpha\alpha}(\phi) + \tilde{O}_3^{\alpha\alpha\beta}(\phi) \\ &\quad\quad+\tilde{O}_3^{\alpha\beta\beta}(\phi) +\tilde{O}_3^{\beta\beta\beta}(\phi),
\end{split} \label{eq:O3}
\end{align}
where
\begin{align}
    \tilde{O}_3^{\alpha\alpha\alpha}(\phi) &= \frac{1}{6} (C_{\alpha\alpha\alpha})_{ijk}^{abc} G_{ai}^\alpha G_{bj}^\alpha G_{ck}^\alpha \\
    \tilde{O}_3^{\alpha\alpha\beta}(\phi) &= \frac{1}{2} (C_{\alpha\alpha\beta})_{ijk}^{abc} G_{ai}^\alpha G_{bj}^\alpha G_{ck}^\beta \\
    \tilde{O}_3^{\alpha\beta\beta}(\phi) &= \frac{1}{2} (C_{\alpha\beta\beta})_{ijk}^{abc} G_{ai}^\alpha G_{bj}^\beta G_{ck}^\beta \\
    \tilde{O}_3^{\beta\beta\beta}(\phi) &= \frac{1}{6} (C_{\beta\beta\beta})_{ijk}^{abc} G_{ai}^\beta G_{bj}^\beta G_{ck}^\beta. \label{eq:last-O3}
\end{align}
These expressions can be derived from the generalized Wick's theorem.
Considering $\tilde{O}_3^{\alpha\alpha\alpha}(\phi)$ in detail, we have
\begin{align}
\begin{split}
    \tilde{O}_3^{\alpha\alpha\alpha}(\phi) &= \frac{1}{36} (C_{\alpha\alpha\alpha})_{ijk}^{abc} \frac{\mel*{\phi_0}{\cre{a\alpha} \cre{b\alpha} \cre{c\alpha} \ani{k\alpha} \ani{j\alpha} \ani{i\alpha}}{\phi}}{\braket{\phi_0}{\phi}} \\
    &= \frac{1}{36} (C_{\alpha\alpha\alpha})_{ijk}^{abc} \begin{vmatrix}
        G_{ai}^\alpha & G_{aj}^\alpha & G_{ak}^\alpha \\
        G_{bi}^\alpha & G_{bj}^\alpha & G_{bk}^\alpha \\
        G_{ci}^\alpha & G_{cj}^\alpha & G_{ck}^\alpha 
    \end{vmatrix} \\
    &= \frac{1}{6} (C_{\alpha\alpha\alpha})_{ijk}^{abc} G_{ai}^\alpha G_{bj}^\alpha G_{ck}^\alpha,
\end{split} \label{eq:first-O3}
\end{align}
 where we have adopted the Einstein summation convention and made use of  symmetries in the $C_{\alpha\alpha\alpha}$ tensor to identify terms that are identical. 

\subsubsection*{Implementation details}
For trials with both triple and quadruple excitations, we have implemented $O(\phi)$ in the automatic differentiation framework in \adafqmc.\citep{ad_afqmc_software} For trials with triple excitations, we have furthermore implemented $E_L(\phi)$ directly to reduce computational costs; see Appendix \ref{app:ccsdt-manual} for details. The implementations were tested against auto-generated code for overlaps and local energies with arbitrary $\vert \phi \rangle$ for CISDT and CISDTQ trial states. All of our AFQMC implementations support GPU acceleration via \textsc{jax}.\citep{bradbury2018jax} Since one needs to compute the trial state before the AFQMC calculation is performed, the overall computational scalings of the AFQMC/CCSDT and AFQMC/CCSDTQ implementations are the same as that to determine the CCSDT and CCSDTQ trial states, i.e., $N^8$ and $N^{10}$, respectively, where $N$ denotes the size of the system. However, neglecting the cost to determine the trial, the scaling of the AFQMC calculation is (in an optimal implementation) $N^7$ and $N^9$ respectively, for a fixed stochastic error. 
 For more details regarding computational scaling, see Appendix \ref{app:scaling}. Finally, to determine the CCSDT and CCSDTQ trials, we have interfaced \adafqmc~ with \textsc{ccpy}.\citep{gururangan2024ccpy} 

\section{Results and discussion} \label{sec:results-discussion}

We benchmark the AFQMC/CC 
methods on a subset of the HEAT dataset, a set of small molecules at ground state equilibrium geometries.\citep{tajti2004heat}
In these calculations, we use the 6-31G basis and a restricted/unrestricted Hartree-Fock reference for closed/open-shell systems. We obtain the high order coupled cluster trials by using \textsc{ccpy}. Reference energies with pentuple (CCSDTQP) and hextuple (CCSDTQPH) excitations were computed with \textsc{mrcc}.\citep{kallay2020mrcc} The walkers are spin-restricted and we have frozen the core orbitals for the closed-shell systems.
The AFQMC calculations are performed with 400 walkers propagated for 2000 blocks, with 50 propagation steps for each block, and a timestep of $0.005$ a.u. The Cholesky decomposition threshold for the electron repulsion integrals was set to $10^{-8}$ a.u.
Input files and output data for all calculations are available in a public repository,\citep{zenodo} and the code for trials with triple and quadruple excitations is available in \adafqmc.
Throughout this work, we have performed the \adafqmc \ calculations using NVIDIA A100 GPUs on the Perlmutter supercomputer. 

\begin{figure*}[htb]
    \centering
    \includegraphics[width=\linewidth]{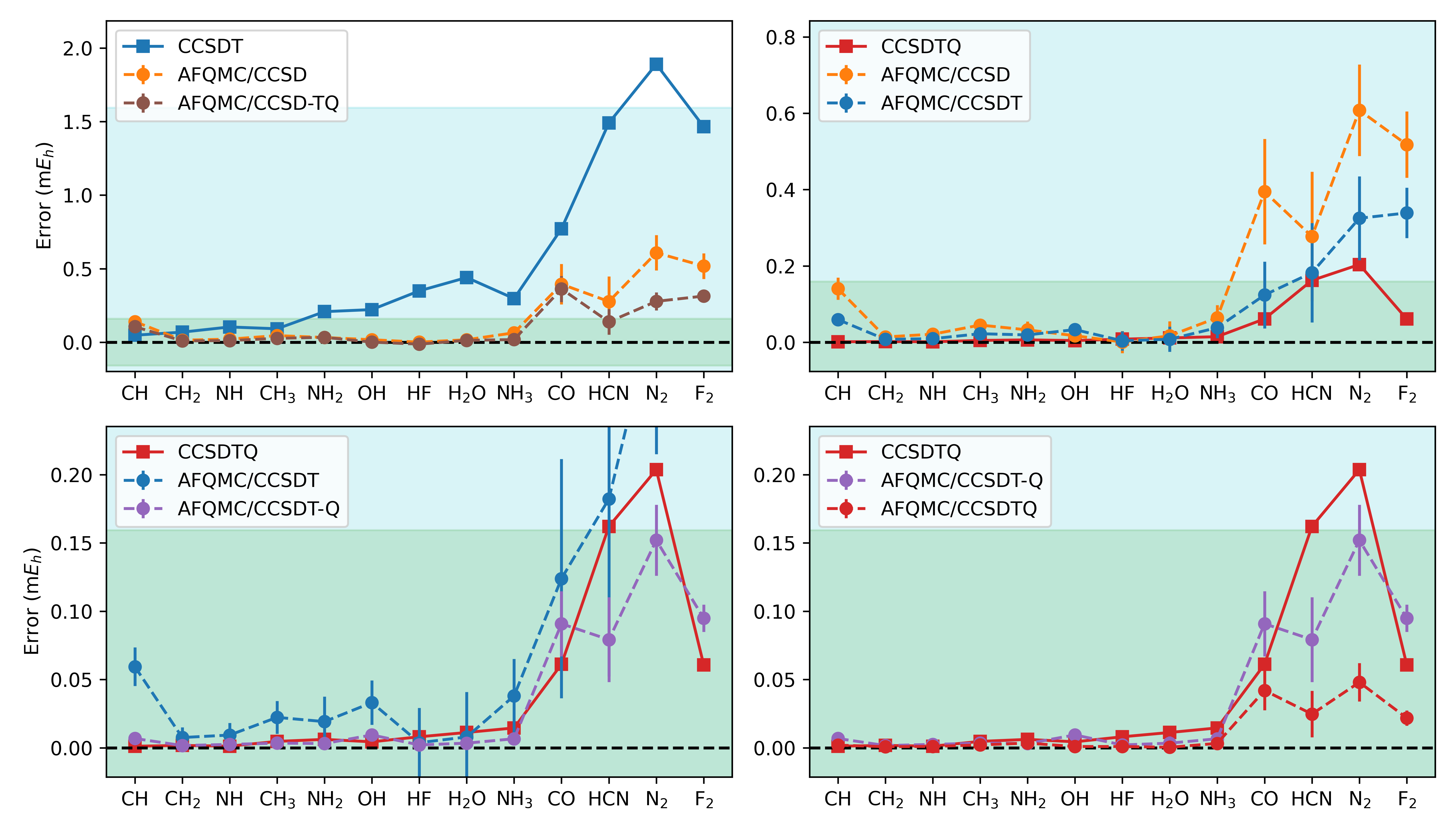}
    \caption{Errors in the ground state energy for the HEAT dataset. Shaded areas indicate chemical accuracy $E_{ca} = 1.596$ m$E_h$ $\approx 1$ kcal/mol (blue) and 0.1 kcal/mol (green). Errors are given relative to CCSDTQP (for F$_2$) and CCSDTQPH (for all other systems).}
    \label{fig:heat}
\end{figure*}

Errors in the ground state energies are shown in Fig.~\ref{fig:heat} and average errors are given in Table \ref{tab:heat}. Starting with AFQMC/CCSD (Fig.~\ref{fig:heat}, top left), we find errors are generally within chemical accuracy, consistent with previous calculations on the same dataset using the cc-pVDZ basis set.\citep{mahajan2025beyond} In nearly all cases, the AFQMC/CCSD energies outperform CCSDT, and in all cases, the errors are significantly improved compared to the CCSD trial. Moreover, we find that the energies are improved by approximately a factor of two when including disconnected triple and quadruple excitations (denoted CCSD-TQ) in the trial.

\begin{table}[htb]
    \centering
    \caption{Average energy errors (in m$E_h$) for the trial ($\bar{\Delta}_T$) and the AFQMC estimate ($\bar{\Delta}_\mathrm{AFQMC}$) for the HEAT dataset. }
    \begin{ruledtabular}
    \begin{tabular}{lccc}
           & $\bar{\Delta}_T$ & $\bar{\Delta}_\mathrm{AFQMC}$ & $\bar{\Delta}_T/\bar{\Delta}_\mathrm{AFQMC}$ \\
           \midrule
     HF    & 142.05 & 1.25 & 113.7  \\
     CCSD  & 3.43 & 0.17 & 20.8 \\
     CCSDT & 0.57 & 0.09 & 6.4 \\
     CCSDTQ & 0.04 & 0.01 & 3.6 \\ 
     \midrule
     CCSD-TQ & 3.43 & 0.10 & 34.4 \\
     CCSDT-Q & 0.57 & 0.04 & 16.3 
    \end{tabular}
    \end{ruledtabular}
    \label{tab:heat}
\end{table}

The improvements in the energies relative to the trial are smaller for AFQMC/CCSDT (Fig.~\ref{fig:heat}, top right). While the AFQMC/CCSDT energies are generally more accurate than CCSDT, the errors are sometimes comparable to AFQMC/CCSD and are much less accurate than CCSDTQ. However, the AFQMC/CCSDT energies can be significantly improved by including disconnected quadruple excitations in the CCSDT trial.
This AFQMC/CCSDT-Q approach gives errors similar to CCSDTQ for the HEAT dataset (see Fig.~\ref{fig:heat}, bottom left).  
To perform the calculations with disconnected triple and quadruple excitations, we have used the $N^{10}$ AFQMC/CCSDTQ implementation. 
The disconnected nature of these terms can in principle be exploited to lower the computational costs. In particular, AFQMC/CCSDT-Q can be implemented with the same $N^8$ scaling as AFQMC/CCSDT by not explicitly forming the quadruples tensor, albeit at the cost of a higher prefactor. 
Finally, for AFQMC/CCSDTQ, we find improvements beyond CCSDTQ (Fig.~\ref{fig:heat}, bottom right). 

Overall, the results show that the error is systematically improved by increasing the trial quality. We also see that (in these systems) the AFQMC energy  improves on the trial CC energy, and because the CC series converges rapidly, so does the AFQMC/CC error. However, the improvement relative to the trial is most dramatic for the least accurate trials,  as can be seen from the relative improvements listed in Table \ref{tab:heat}.
In practice, in the small systems and basis sets studied here and for our chosen computational setup, the determination of the trial takes less walltime than the subsequent AFQMC calculation, e.g.~for \ce{N2}, the walltimes (using 32 cores on an AMD EPYC 7763 CPU for CC, and 1 Nvidia A100 GPU for AFQMC/CC) were CCSDT: 1.5~s,  AFQMC/CCSDT: 8 minutes 46 s; CCSDTQ: 4 minutes 30 s; AFQMC/CCSDTQ: 14 hours. However, since the  scaling of the AFQMC step is lower than that of computing the trial, in larger systems the improvement in accuracy obtained by applying AFQMC may still be justified when using more accurate CC trials.

\begin{figure*}[htb]
    \centering
    \includegraphics[width=\linewidth]{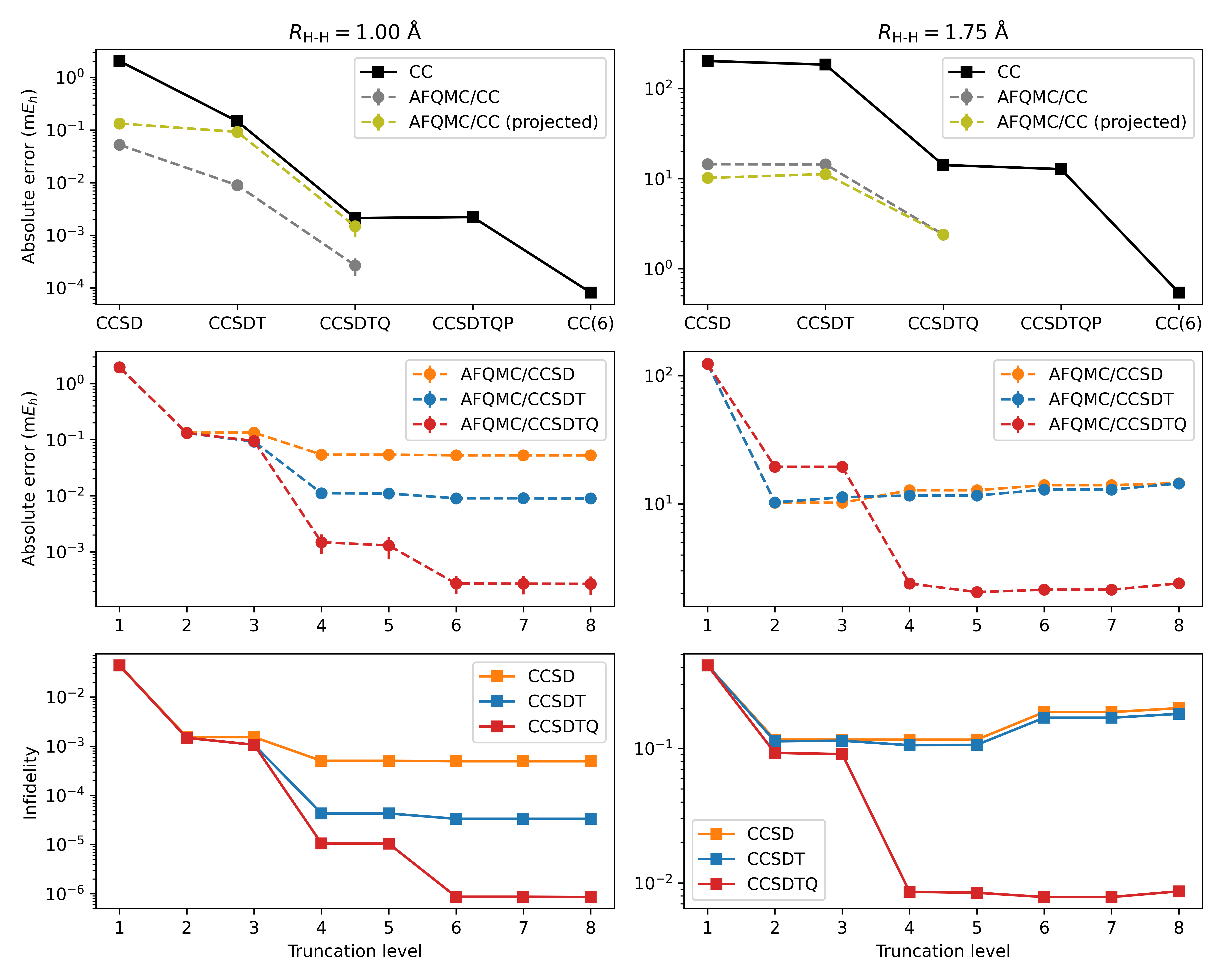}
    \caption{Hydrogen cluster (H$_8$/STO-3G). The left and right columns show results for H--H bond distances of $R = 1.00$ Å and $R = 1.75$ Å. The upper panels show errors in energies for fully expanded and projected trials, where the truncation of the projected trials is given as the truncation of the cluster operator. The middle panels show absolute errors for varying truncations of the trial. The bottom panels show the infidelity of the truncated trials (see text).}
    \label{fig:h8}
\end{figure*}

The above results revealed an important role of the disconnected contributions in the trial wavefunction at equilibrium geometries.
To further investigate the impact of the CI projection of the CC wavefunction, we consider a cubic hydrogen cluster (H$_8$/STO-3G) for which we can systematically consider the trials
\begin{align}
    \vert \psi_T^{k,n} \rangle =  \Bigl( \sum_{j=0}^k \mathcal{P}_j \Bigr) e^{\sum_{i=1}^n T_i} \vert \phi_0 \rangle,
\end{align}
with $k$ and $n$ being the truncation levels of $\mathcal{P}$ and $T$. To apply these trials, we convert the coupled cluster state into its CI expansion, zeroing out terms beyond the excitation level $k$, and use the resulting expansion in the AFQMC/MSD implementation in \adafqmc. For these calculations, we have used a restricted Hartree-Fock reference. The Cholesky threshold was set to $10^{-10}$ a.u.~and we use the same number of walkers and blocks as for the HEAT dataset (400 walkers, 2000 blocks, 50 steps per block, and a timestep of 0.005 a.u.).

\begin{table*}[htb]
    \centering
    \caption{Error fraction $\Delta_{\PsiCC}/\Delta_{\mathrm{AFQMC}}$ for H$_8$/STO-3G. Values are given for both coupled cluster trials truncated to the CI subspace (denoted proj) and with no projection of the coupled cluster state (denoted full).}
    \begin{ruledtabular}
    \begin{tabular}{lcccc}
     & \multicolumn{2}{c}{$R = 1.00$ Å} & \multicolumn{2}{c}{$R = 1.75$ Å} \\
     \cmidrule{2-3}\cmidrule{4-5}
      & $\Delta_{\PsiCC}/\Delta_{\mathrm{AFQMC}}^\mathrm{proj}$ & $\Delta_{\PsiCC}/\Delta_{\mathrm{AFQMC}}^\mathrm{full}$  & $\Delta_{\PsiCC}/\Delta_{\mathrm{AFQMC}}^\mathrm{proj}$ & $\Delta_{\PsiCC}/\Delta_{\mathrm{AFQMC}}^\mathrm{full}$  \\
      \midrule  
     CCSD    & 15.4 & 39.2 & 19.9 & 14.0\\
     CCSDT   & 1.6  & 16.5 & 16.4 & 12.8\\
     CCSDTQ  & 1.4  & 7.9  & 5.9  & 5.9 \\
    \end{tabular}
    \end{ruledtabular}
    \label{tab:h8-relative-improvements}
\end{table*}

The results on H$_8$ are summarized in Fig.~\ref{fig:h8}. We consider the performance of the trials in the weakly and strongly correlated regimes ($R=1.00$ Å, $R=1.75$ Å). Starting with the upper left and right panels, we see that using the CC wavefunction trial with no CI projection (gray curve) leads to significant improvements in the energies, relative to using CI projection, for the weakly correlated case (left) but not for the strongly correlated case (right). To discern the importance of individual excitation orders, we compute the errors obtained using different projection levels for the trial in the middle left and right panels. In the weakly correlated case (left), we see that quadruple excitations improve the AFQMC/CCSD and AFQMC/CCSDT energies and that hextuple excitations improve the AFQMC/CCSDTQ energies. In contrast, in the strongly correlated case (right), higher-order excitations do not improve the energies and can even make the energies less accurate (see values for CCSD and CCSDT).  

The effectiveness of including higher-order contributions is related to whether their inclusion improves the trial, as we can show by computing the trial infidelity
\begin{align}
\begin{split}
    f^{k,n} &= 1 - O^{k,n},
\end{split}
\end{align}
where
\begin{align}
    O^{k,n} = \frac{\braket*{\psi_T^{k,n}}{\psi_0}}{\braket*{\psi_T^{k,n}}^{1/2} \braket*{\psi_0}^{1/2}}.
\end{align}
Here, $\vert \psi_0 \rangle$ is the full configuration interaction wave function.
Comparing the middle and lower panels of Fig.~\ref{fig:h8}, where we present $f^{k,n}$ for projection levels $k=1,2,\ldots,8$, we see that reduced errors coincide with improved trial quality as measured by $f^{k,n}$. We emphasize that this should not be taken to mean that increased fidelity of the trial necessarily lowers the error. For example, for $R = 1.75$ Å and truncation level $k = 2$, the projected CCSDTQ trial has a higher fidelity compared with the CCSD and CCSDT trials, but a larger error in the energy (see middle and bottom right panels).

Like for the HEAT dataset, we observe a pattern of diminishing returns in H$_8$ as we step up the AFQMC/CC hierarchy (see Table \ref{tab:h8-relative-improvements}). Note that this pattern is present both for the projected and the non-projected trials.

\begin{figure}[htb]
    \centering
    \includegraphics[width=\linewidth]{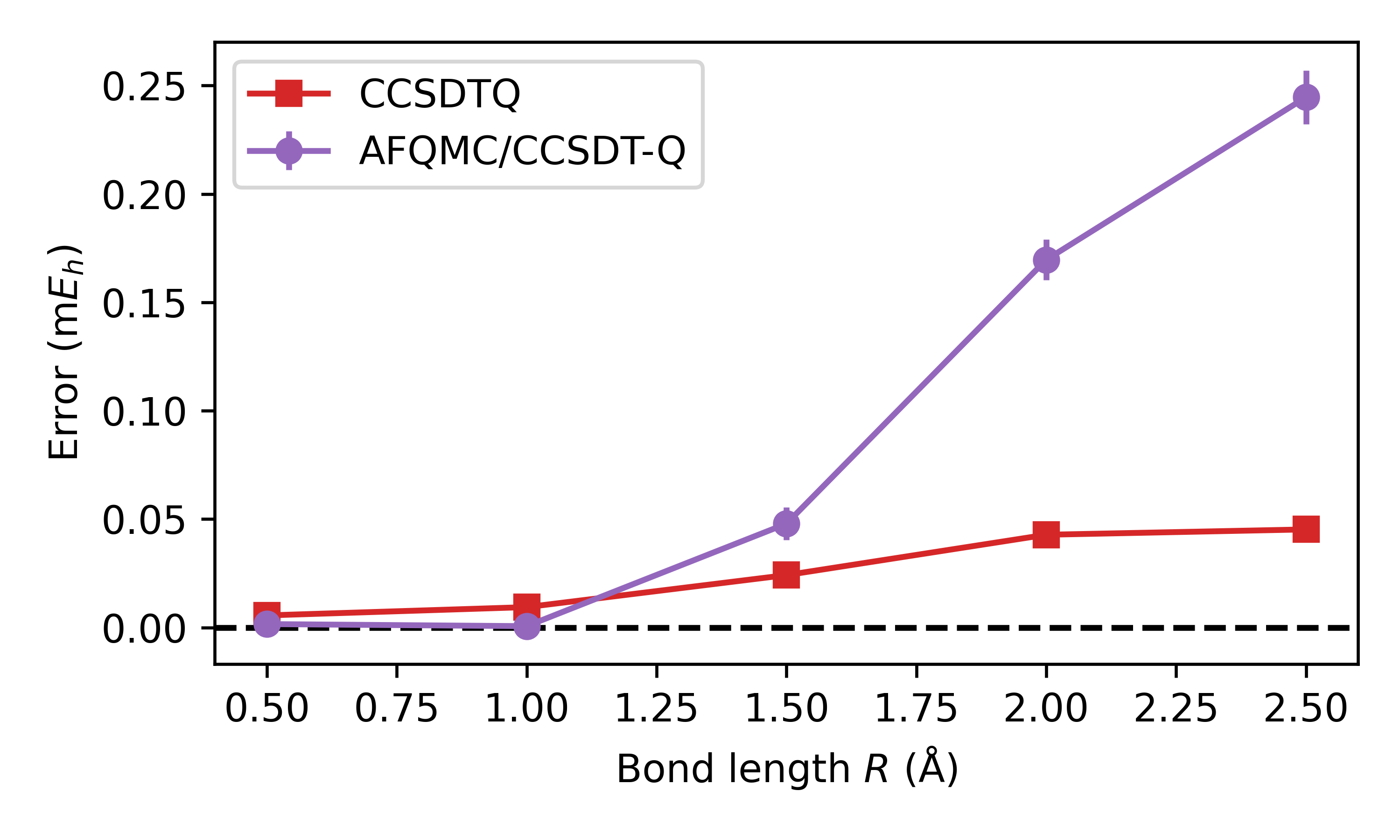}
    \caption{Errors along potential energy curve for HF/6-31G.}
    \label{fig:hf-dissociation}
\end{figure}

The reduced errors obtained by including quadruples in the CCSDT trial for $R = 1.00$ Å in H$_8$ mirrors the improvements for CCSDT-Q on the HEAT dataset (see Table \ref{tab:heat}). However, the results for $R = 1.75$ Å suggest that these improvements are limited to regions close to equilibrium. Consistent with this result, we find for the HF molecule that the AFQMC/CCSDT-Q error increases relative to CCSDTQ for longer bond lengths (see Fig.~\ref{fig:hf-dissociation}). 

\section{Conclusions} \label{sec:conclusions}
In this work, we have investigated the systematic improvement of trial states in phaseless AFQMC. The use of CCSD trial states, projected onto a CI subspace, was recently found to provide promising results on molecular systems, often outperforming CCSD(T).\citep{mahajan2025beyond} Prompted by this finding, we extended the implementation to include coupled cluster trials with triple (CCSDT) and quadruple (CCSDTQ) excitations.

Our benchmark results on the HEAT dataset and some small bond-breaking examples shows that in these systems, going to CC trials beyond CCSD does improve the error of phaseless AFQMC. On the other hand, the relative improvement is modest compared to the improvement seen when using a CCSD trial. Although in our benchmarks the cost of the AFQMC step was large compared to the CC trial determination, the computational scaling of AFQMC/CC is less than that to compute the trial. Thus employing high order CC trials in AFQMC/CC may be useful in high accuracy calculations of larger problems beyond benchmarking.

We also examined the role of different components of the trial wavefunction in controlling the AFQMC bias. We found that including disconnected contributions of the CC wavefunction in the trial state significantly improves the AFQMC energy for weakly correlated systems. However, for strongly correlated systems, this was not the case, and we even observed that improvements in fidelity of the trial state with the exact ground-state did not always correlate with improved AFQMC energies. 
This points to a need for more research into the relationship between the quality of the trial and the accuracy of phaseless AFQMC. Such work would seem to be of particular relevance to the ongoing efforts into the use of various types of trial states\citep{al2006auxiliary,purwanto2008eliminating,wouters2014projector,jiang2025unbiasing,purwanto2009excited,landinez2019non,mahajan2022selected,sukurma2025self} as well as the more recent efforts into applying the phaseless AFQMC method in hybrid quantum-classical algorithms.\citep{huggins2022unbiasing,amsler2023classical,kiser2024classical,huang2024evaluating,kiser2025contextual,jiang2025walking}

\section*{Acknowledgments}
We thank Dr.~Huanchen Zhai for providing the auto-generated code for overlaps and local energies with arbitrary $\vert \phi \rangle$ for CISDT and CISDTQ trial states, which we used to check the correctness of our manual implementations, and Dr.~Ankit Mahajan for assistance with the \adafqmc~program. Work by EK and GKC was supported by the US Department of Energy, Office of Science, via award no. DE-SC0018140C.
This work used the Delta system at the National Center for Supercomputing Applications through allocation CHE250023 from the Advanced Cyberinfrastructure Coordination Ecosystem: Services \& Support (ACCESS) program, which is supported by U.S. National Science Foundation grants \#2138259, \#2138286, \#2138307, \#2137603, and \#2138296. SS and YD were supported by the DOE grant DE-SC0025943. 

\appendix 
\section{Quadruples implementation of overlap} \label{app:ccsdtq}
The overlap-based implementation for quadruples follows the same recipe outlined in Section \ref{sec:ccsdt-overlap-impl}. The cluster operator for quadruples can be expressed as
\begin{align}
\begin{split}
    T_4 &= \frac{1}{(4!)^2} T_4^{\alpha\alpha\alpha\alpha} + \frac{\binom{4}{1}^2}{(4!)^2}T_4^{\alpha\alpha\alpha\beta} \\
    &+ \frac{\binom{4}{2}^2}{(4!)^2}T_4^{\alpha\alpha\beta\beta} + \frac{\binom{4}{3}^2}{(4!)^2}T_4^{\alpha\beta\beta\beta} + \frac{1}{(4!)^2} T_4^{\beta\beta\beta\beta}
\end{split} \label{eq:T4-unrestricted}
\end{align}
and similarly for $C_4$. The spin blocks of $C_4$ are identified by making use of the identity
\begin{align}
    C_4 = T_4 + T_1 T_3 + \frac{1}{2} T_2^2 + \frac{1}{2} T_1^2 T_2 + \frac{1}{24} T_1^4
\end{align}
and substituting the spin-unrestricted expressions for $T_1$, $T_2$, $T_3$, and $T_4$ given in Eqs.~\eqref{eq:T-unrestricted} and \eqref{eq:T4-unrestricted}. The resulting spin-blocks are
\begin{align}
\begin{split}
    C_4^{\alpha\alpha\alpha\alpha} &= T_4^{\alpha\alpha\alpha\alpha} + 16 T_1^\alpha T_3^{\alpha\alpha\alpha} + 18 (T_2^{\alpha\alpha})^2 \\
    &+ 72 (T_1^\alpha)^2 T_2^{\alpha\alpha} + 24 (T_1^\alpha)^4
\end{split} \\
\begin{split}
    C_4^{\alpha\alpha\alpha\beta} &= T_4^{\alpha\alpha\alpha\beta} + T_3^{\alpha\alpha\alpha} T_1^\beta + 9 T_3^{\alpha\alpha\beta} T_1^\alpha \\
    &+ 9 T_2^{\alpha\alpha} T_2^{\alpha\beta} + 18 (T_1^\alpha)^2 T_2^{\alpha\beta} + 9 T_2^{\alpha\alpha} T_1^\alpha T_1^\beta \\
    &+ 6 (T_1^\alpha)^3 T_1^\beta
\end{split} \\
\begin{split}
    C_4^{\alpha\alpha\beta\beta} &= T_4^{\alpha\alpha\beta\beta} + 4 T_3^{\alpha\alpha\beta} T_1^\beta + 4 T_3^{\alpha\beta\beta} T_1^\alpha \\
    &+ T_2^{\alpha\alpha} T_2^{\beta\beta} + 8 T_2^{\alpha\beta}T_2^{\alpha\beta} + 2 (T_1^\alpha)^2 T_2^{\beta\beta} \\
    &+ 2 T_2^{\alpha\alpha} (T_1^\beta)^2 + 16 T_1^\alpha T_1^\beta T_2^{\alpha\beta} \\
    &+ 4 (T_1^\alpha)^2 (T_1^\beta)^2
\end{split} \\
\begin{split}
    C_4^{\alpha\beta\beta\beta} &= T_4^{\alpha\beta\beta\beta} + 9 T_3^{\alpha\beta\beta} T_1^\beta + T_3^{\beta\beta\beta} T_1^\alpha \\
    &+ 9 T_2^{\alpha\beta} T_2^{\beta\beta} + 18 (T_1^\beta)^2 T_2^{\alpha\beta} + 9 T_1^\alpha T_1^\beta T_2^{\beta\beta} \\
    &+ 6 (T_1^\beta)^3 T_1^\alpha
\end{split} \\
\begin{split}
    C_4^{\beta\beta\beta\beta} &= T_4^{\beta\beta\beta\beta} + 16 T_1^\beta T_3^{\beta\beta\beta} + 18 (T_2^{\beta\beta})^2 \\
    &+ 72 (T_1^\beta)^2 T_2^{\beta\beta} + 24 (T_1^\beta)^4
\end{split} 
\end{align}
As for the triples, the tensors in each quadruples block are symmetrized with respect to exchange of excitations  ($ai \leftrightarrow bj$) and anti-symmetrized  with respect to exchanging occupied ($i \leftrightarrow j$) and virtual ($a \leftrightarrow b$) orbitals.

The generalized Wick's theorem implies that
\begin{align}
\begin{split}
    \tilde{O}_4(\phi) &= \tilde{O}_4^{\alpha\alpha\alpha\alpha}(\phi) + \tilde{O}_4^{\alpha\alpha\alpha\beta}(\phi) + \tilde{O}_4^{\alpha\alpha\beta\beta}(\phi) \\
    &+ \tilde{O}_4^{\alpha\beta\beta\beta}(\phi) + \tilde{O}_4^{\beta\beta\beta\beta}(\phi) 
\end{split}
\end{align}
where
\begin{align}
    \tilde{O}_4^{\alpha\alpha\alpha\alpha}(\phi) &= \frac{1}{24} (C_{\alpha\alpha\alpha\alpha})_{ijkl}^{abcd} G_{ai}^\alpha G_{bj}^\alpha G_{ck}^\alpha G_{dl}^\alpha \\
    \tilde{O}_4^{\alpha\alpha\alpha\beta}(\phi) &= \frac{1}{6} (C_{\alpha\alpha\alpha\beta})_{ijkl}^{abcd} G_{ai}^\alpha G_{bj}^\alpha G_{ck}^\alpha G_{dl}^\beta \\
    \tilde{O}_4^{\alpha\alpha\beta\beta}(\phi) &= \frac{1}{4} (C_{\alpha\alpha\beta\beta})_{ijkl}^{abcd} G_{ai}^\alpha G_{bj}^\alpha G_{ck}^\beta G_{dl}^\beta \\
    \tilde{O}_4^{\alpha\beta\beta\beta}(\phi) &= \frac{1}{6} (C_{\alpha\beta\beta\beta})_{ijkl}^{abcd} G_{ai}^\alpha G_{bj}^\beta G_{ck}^\beta G_{dl}^\beta \\
    \tilde{O}_4^{\beta\beta\beta\beta}(\phi) &= \frac{1}{24} (C_{\beta\beta\beta\beta})_{ijkl}^{abcd} G_{ai}^\beta G_{bj}^\beta G_{ck}^\beta G_{dl}^\beta.
\end{align}
These terms are derived similarly as described for the triples. For example, we have
\begin{align}
\begin{split}
    \tilde{O}_4^{\alpha\alpha\beta\beta}(\phi) &= \frac{\binom{4}{2}^2}{(4!)^2} \frac{\mel*{\phi_0}{C_4^{\alpha\alpha\beta\beta}}{\phi}}{\braket{\phi_0}{\phi}} \\
    &=  \frac{\binom{4}{2}^2}{(4!)^2} (C_{\alpha\alpha\beta\beta})_{ijkl}^{abcd} \begin{vmatrix}
        G_{ai}^\alpha & G_{aj}^\alpha & 0 & 0 \\
        G_{bi}^\alpha & G_{bj}^\alpha & 0 & 0 \\
        0 & 0 & G_{ck}^\beta & G_{cl}^\beta \\
        0 & 0 & G_{dk}^\beta & G_{dl}^\beta
    \end{vmatrix} \\
    &= \frac{\binom{4}{2}^2}{(4!)^2} (2!)^2 (C_{\alpha\alpha\beta\beta})_{ijkl}^{abcd} G_{ai}^\alpha G_{bj}^\alpha G_{ck}^\beta G_{dl}^\beta \\
    &= \frac{1}{4} (C_{\alpha\alpha\beta\beta})_{ijkl}^{abcd} G_{ai}^\alpha G_{bj}^\alpha G_{ck}^\beta G_{dl}^\beta,
\end{split}
\end{align}
where identical terms from the determinant provide an overall pre-factor for the contraction of $C_{\alpha\alpha\beta\beta}$ with the Green's functions. 

\section{Triples implementation of force bias and local energy} \label{app:ccsdt-manual}
We group the force bias and local energy contributions arising from different spin blocks and apply the non-orthogonal Wick's theorem. 
The terms below are restricted to the contributions due to $C_3$. 
We use the index convention in Ref.~\citenum{mahajan2025beyond}, denoting general molecular orbitals by $i,j$, occupied orbitals by $p,q,r$, and virtual orbitals by $t,u,s$.   
Programmable expressions for terms due to $C_1$ and $C_2$ can be found in Ref.~\citenum{mahajan2025beyond}. We adopt Einstein summation to keep the notation simple. 
Like for the overlap, we define scaled quantities $\tilde{X}$ that are amenable to the application of Wick's theorem, i.e.~we provide expressions for $\tilde{X}$ where $X = \tilde{X} O_0$ is the quantity of interest. 

For the force bias, we have
\begin{align}
    \begin{split}
        \tilde{v}_\gamma^{\alpha\alpha\alpha} &= 
         \frac{1}{6} c_{pqr}^{tus}  \Bigl( \bigl( L_{ij}^{\gamma,\sigma} G_{ij}^\sigma \bigr) G_{pt}^\alpha G_{qu}^\alpha G_{rs}^\alpha - 3 L_{ij}^{\gamma,\alpha} \mathcal{G}_{it}^\alpha G_{pj}^\alpha G_{qu}^\alpha G_{rs}^\alpha \Bigr),
    \end{split}
\end{align}
where $\mathcal{G}_{it}^\alpha = G_{it}^\alpha - \delta_{it}$.
The expression for $\tilde{v}_\gamma^{\beta\beta\beta}$ is given by $\alpha \leftrightarrow \beta$. Here, and in the following, it is to be understood that $c_{pqr}^{tus}$ refers to the spin-block of $C_3$ in question (here, $c \equiv C^{\alpha\alpha\alpha}$). The mixed-spin blocks are given as:
\begin{align}
\begin{split}
    \tilde{v}_\gamma^{\alpha\alpha\beta} 
    &= \frac{1}{2} c_{pqr}^{tus} \Bigl( \bigl( L_{ij}^{\gamma,\sigma} G_{ij}^\sigma \bigr) G_{pt}^\alpha G_{qu}^\alpha G_{rs}^\beta - 2 L_{ij}^{\gamma,\alpha} \mathcal{G}_{it}^\alpha G_{pj}^\alpha G_{qu}^\alpha G_{rs}^\beta - L_{ij}^{\gamma,\beta} \mathcal{G}_{is}^\beta G_{rj}^\beta G_{pt}^\alpha G_{qu}^\alpha \Bigr)
\end{split}
\end{align}
The expression for $\tilde{v}^{\alpha\beta\beta}$ is given by $\alpha \leftrightarrow \beta$. The corresponding terms for the one-electron energy are given by replacing $L_{ij}^\gamma$ by $h_{ij}$. 

For the two-electron energy, we have
\begin{align}
    \tilde{E}_{\alpha\alpha\alpha}^{(1)} 
    &= \frac{1}{12} L_{ij}^{\gamma,\sigma} G_{ij}^{\sigma} c_{pqr}^{tus} ( L_{kl}^{\gamma,\lambda} G_{kl}^{\lambda} G_{pt}^\alpha G_{qu}^\alpha G_{rs}^\alpha  - 3 L_{kl}^{\gamma,\alpha} \mathcal{G}_{kt}^\alpha G_{pl}^\alpha G_{qu}^\alpha G_{rs}^\alpha ) \\ 
    \tilde{E}_{\alpha\alpha\alpha}^{(2)} &= - \frac{1}{12}  c_{pqr}^{tus} \Bigl( \bigl( L_{ij}^{\gamma,\sigma} L_{kl}^{\gamma,\sigma} G_{il}^\sigma G_{kj}^\sigma \bigr) G_{pt}^\alpha G_{qu}^\alpha G_{rs}^\alpha - 3 L_{ij}^{\gamma,\alpha} L_{kl}^{\gamma,\alpha}  G_{il}^\alpha \mathcal{G}_{kt}^\alpha G_{pj}^\alpha G_{qu}^\alpha G_{rs}^\alpha \Bigr) \\
    \tilde{E}_{\alpha\alpha\alpha}^{(3)} &= \frac{1}{4} L_{ij}^{\gamma,\alpha}  c_{pqr}^{tus} \mathcal{G}_{it}^\alpha \bigl(  L_{kl}^{\gamma,\alpha}  G_{kj}^\alpha G_{pl}^\alpha G_{qu}^\alpha G_{rs}^\alpha + 2 L_{kl}^{\gamma,\alpha} \mathcal{G}_{ku}^\alpha G_{pj}^\alpha G_{ql}^\alpha G_{rs}^\alpha -  L_{kl}^{\gamma,\lambda} G_{kl}^{\lambda} G_{pj}^\alpha G_{qu}^\alpha G_{rs}^\alpha \bigr) 
\end{align}
and
\begin{align}
    \begin{split}
        \tilde{E}_{\alpha\alpha\beta}^{(1)} &= \frac{1}{4} L_{ij}^{\gamma,\sigma}G_{ij}^{\sigma} c_{pqr}^{tus} ( L_{kl}^{\gamma,\lambda} G_{kl}^{\lambda}  G_{pt}^\alpha G_{qu}^\alpha G_{rs}^\beta - 2 L_{kl}^{\gamma, \alpha} \mathcal{G}_{kt}^\alpha G_{pl}^\alpha G_{qu}^\alpha G_{rs}^\beta 
                                - L_{kl}^{\gamma,\beta} \mathcal{G}_{ks}^\beta G_{pt}^\alpha G_{qu}^\alpha G_{rl}^\beta) 
    \end{split} \\
    \begin{split}
        \tilde{E}_{\alpha\alpha\beta}^{(2)} &= -\frac{1}{4} c_{pqr}^{tus}( L_{ij}^{\gamma,\sigma}L_{kl}^{\gamma,\sigma} G_{il}^{\sigma} G_{kj}^{\sigma} G_{pt}^\alpha G_{qu}^\alpha G_{rs}^\beta - 2 L_{ij}^{\gamma,\alpha}L_{kl}^{\gamma,\alpha}G_{il}^{\alpha} \mathcal{G}_{kt}^\alpha G_{pj}^\alpha G_{qu}^\alpha G_{rs}^\beta \\
                                &\quad- L_{ij}^{\gamma,\beta}L_{kl}^{\gamma,\beta}G_{il}^{\beta} \mathcal{G}_{ks}^\beta G_{rj}^\beta G_{pt}^\alpha G_{qu}^\alpha) 
    \end{split} \\
    \begin{split}
        \tilde{E}_{\alpha\alpha\beta}^{(3)} &= \frac{1}{2} L_{ij}^{\gamma,\alpha} \mathcal{G}_{it}^\alpha c_{pqr}^{tus} (L_{kl}^{\gamma,\alpha} G_{kj}^\alpha G_{pl}^\alpha G_{qu}^\alpha G_{rs}^\beta - L_{kl}^{\gamma,\lambda} G_{kl}^{\lambda} G_{pj}^\alpha G_{qu}^\alpha G_{rs}^\beta \\
    &\quad+ L_{kl}^{\gamma,\alpha} \mathcal{G}_{ku}^\alpha G_{pj}^\alpha G_{ql}^\alpha G_{rs}^\beta + L_{kl}^{\gamma,\beta}\mathcal{G}_{ks}^\beta G_{rl}^\beta G_{pj}^\alpha G_{qu}^\alpha) 
\end{split} \\
\tilde{E}_{\alpha\alpha\beta}^{(4)} &= \frac{1}{4} L_{ij}^{\gamma,\beta} \mathcal{G}_{is}^\beta c_{pqr}^{tus} ( L_{kl}^{\gamma,\beta} G_{kj}^\beta G_{rl}^\beta G_{pt}^\alpha G_{qu}^\alpha - L_{kl}^{\gamma,\lambda}G_{kl}^{\lambda} G_{rj}^\beta G_{pt}^\alpha G_{qu}^\alpha + 2 L_{kl}^{\gamma,\alpha} \mathcal{G}_{kt}^\alpha G_{rj}^\beta G_{pl}^\alpha G_{qu}^\alpha ).
\end{align}
The $\tilde{E}_{\beta\beta\beta}$ and $\tilde{E}_{\alpha\beta\beta}$ blocks are given by $\alpha \leftrightarrow \beta$.

\section{Computational scaling for triple and quadruple excitations} \label{app:scaling}
The computational costs can be subdivided into the cost to determine the trial and that associated with stochastic sampling for the given trial.
To obtain the coupled cluster trial up to order $n$, the cost scales as $N^{2n+2}$, where $N$ is the number of orbitals in the system. In the cases of CCSDT and CCSDTQ, this scaling is therefore $N^8$ and $N^{10}$, respectively.

To determine the sampling cost, we need to consider the cost of evaluating $O(\phi)$, the force bias, and $E_L(\phi)$, as well as the  cost to obtain a fixed stochastic error in the energy. 
 The $O(\phi)$ cost scales as $N^6$ for trials with up to triple excitations and $N^8$ for trials with up to quadruple excitations.  
 The cost of $E_L(\phi)$ differs depending on whether it is implemented directly or obtained via overlap derivatives. The first derivative of $O(\phi)$ has the same scaling as $O(\phi)$, whereas the second derivative scales higher by one order. In an implementation based on overlap derivatives, therefore, $E_L(\phi)$ scales as $N^7$ and $N^9$ for trials with up to triple and quadruple excitations. The scaling of the force bias is the same, since it is obtained as first derivatives calculated for each auxiliary field component.\citep{mahajan2025beyond}
  For a given fixed stochastic error, the scaling is further increased by one order, thus giving an $N^8$ and $N^{10}$ scaling for triples and quadruples---the same scaling as that to determine the trial. 
  
  In contrast, for a \emph{direct} implementation of $E_L(\phi)$, the scaling is $N^6$ and $N^8$ for a triples and quadruples trial, respectively, thereby giving a lower effective scaling of $N^7$ and $N^9$ for a fixed stochastic error. 

\bibliography{paper}

\end{document}